\documentclass{llncs}
\usepackage{makeidx}  
\usepackage{amsmath} 
\usepackage{amssymb} 
\usepackage[colorinlistoftodos]{todonotes} 
\usepackage{cite}
\usepackage{graphicx}
\usepackage{relsize}
\graphicspath{ {./Imgs/} }
\usepackage{array}
\usepackage{multirow}
\usepackage{rotating}
\usepackage[]{algorithm2e}
\usepackage{algpseudocode}
\usepackage{amsmath}
\usepackage{amsfonts}
\usepackage{amssymb}
\usepackage{tikz}
\usepackage{wasysym}
\usepackage{mathtools}
\usepackage{tabularx}
\usepackage{cleveref}
\usepackage[misc]{ifsym}
\begin{document}
\title{Towards Quantifying Neurovascular Resilience}
%
%
\author{
Stefano Moriconi\inst{1} \and Rafael Rehwald\inst{2} \and Maria A. Zuluaga\inst{3} \and H. Rolf J{\"a}ger\inst{2} \and Parashkev Nachev\inst{2} \and S\'ebastien Ourselin\inst{1} \and M. Jorge Cardoso\inst{1} }
\institute{School of Biomedical Engineering and Imaging Sciences, King's College London
\and
Institute of Neurology, University College London
\and
Universidad Nacional de Colombia, Bogot\'{a}, Colombia}
%
\maketitle
\vskip -10pt
\begin{abstract}
Whilst grading neurovascular abnormalities is critical for prompt surgical repair, no statistical markers are currently available for predicting the risk of adverse events, such as stroke, and the overall resilience of a network to vascular complications.
The lack of compact, fast, and scalable simulations with network perturbations impedes the analysis of the vascular resilience to life-threatening conditions, surgical interventions and long-term follow-up.
We introduce a graph-based approach for efficient simulations, which statistically estimates biomarkers from a series of perturbations on the patient-specific vascular network.
Analog-equivalent circuits are derived from clinical angiographies.
Vascular graphs embed mechanical attributes modelling the impedance of a tubular structure with stenosis, tortuosity and complete occlusions.
We evaluate pressure and flow distributions, simulating healthy topologies and abnormal variants with perturbations in key pathological scenarios.
These describe the intrinsic network resilience to pathology, and delineate the underlying cerebrovascular autoregulation mechanisms.
Lastly, a putative graph sampling strategy is devised on the same formulation, to support the topological inference of uncertain neurovascular graphs.
\vskip -10pt
\end{abstract}
\section{Introduction}
\label{Introduction}
Cerebrovascular diseases, such as stroke, are the biggest source of long-term neurological disability in first world countries. In an acute setting, it is necessary to assess the risk of a neurovascular event, and to decide how best to intervene (i.e. thrombectomy vs thrombolysis). These decisions are currently ill-informed and would greatly benefit from supporting statistical and quantitative neurovascular measurements. In preventive care, vascular features and biomarkers could inform patient screening and allow for long-term stroke risk stratification. Despite the incidence and morbidity of cerebrovascular diseases  \cite{mathers2008global}, current studies are often limited to incidental findings in clinical trials. Group-based, quantitative, predictive analysis remains an unexplored research ground.

Previous hemodynamic case-studies \cite{taylor2009open, steinman2003computational, cebral2005characterization, shojima2004magnitude} have provided localised predictive biomarkers of aneurysm formation and rupture, and indices of pathogenesis for occlusive diseases on small portions of the cerebrovascular tree. These studies commonly rely on computationally-intensive fluid dynamics simulations to accurately quantify the blood-flow patterns and vessel wall characteristics, normally disregarding the natural compensation and redundancy mechanisms of the overall network. Recent work in hemodynamic computational fluid dynamics (CFD) have shown prohibitively long simulations for large and complex vascular networks, making the evaluation of how these networks react to perturbations an open challenge \cite{urick2017review,taylor2009open}.
In \cite{ryu2015coupled,chnafa2017improved,onaizah2017model}, artificial physio-pathological equivalents were introduced as a computationally-efficient approximation of CFD analysis. These approximate models condense mechanical tubular features into simplified and compact lumped-parameters, modelling intrinsic whole-network autoregulation mechanisms.
Even though these first-order approximations were found to correlate well with the fully-resolved CFD hemodynamic simulations in cardiac applications \cite{fossan2018optimization}, neurovascular applications are still in their infancy.
Ideally, these models should be fully data-driven, be able to specify phenotypical traits, and generalise to a number of perturbations. 

Leveraging such efficient approximation and by integrating subject-specific vascular graphs from clinical angiographies, we introduce here a scalable and fast simulation framework that statistically estimates functional biomarkers by perturbing vascular topologies.
In detail, we use a spatial graph of the neurovascular network to simulate steady-state blood flow using an analog-equivalent circuit approximation, thus modelling biomechanical lumped-parameters and topological connectivity directly from clinical scans.
Assuming that a latent autoregulation mechanism underlies major brain arteries, we introduce multiple artificial perturbations (e.g. stenosis, tortuosity, occlusions) and simulate how the neurovascular network reacts to these changes.
We then study how different perturbations lead to pressure/flow alterations, and result in downstream changes in vessel wall tension, providing a new metric of neurovascular resilience to different pathological scenarios.
Beyond estimating biomarkers, a putative graph sampling strategy is lastly devised based on the same analog-equivalent formulation.
This supports the topological inference of uncertain redundant vascular network by increasing the sparsity of a fully-connected neurovascular graph, yet preserving its most biologically-plausible set of realisations. 
\section{Methods}
\label{Methods}
\begin{figure}[t]
\vskip -10pt
    \centering
    \begin{tabular}{@{}c@{}c@{\hspace{1em}}|@{\hspace{1em}}ccc@{}}
         \tiny{\textbf{Phantom}}&\tiny{\textbf{Cross-Sectional Snakes}}&\tiny{\textbf{Vessel Impedance}}&\tiny{\textbf{Blood Flow}}&\tiny{\textbf{Pressure Drop}}\\
         \includegraphics[width=0.2\textwidth]{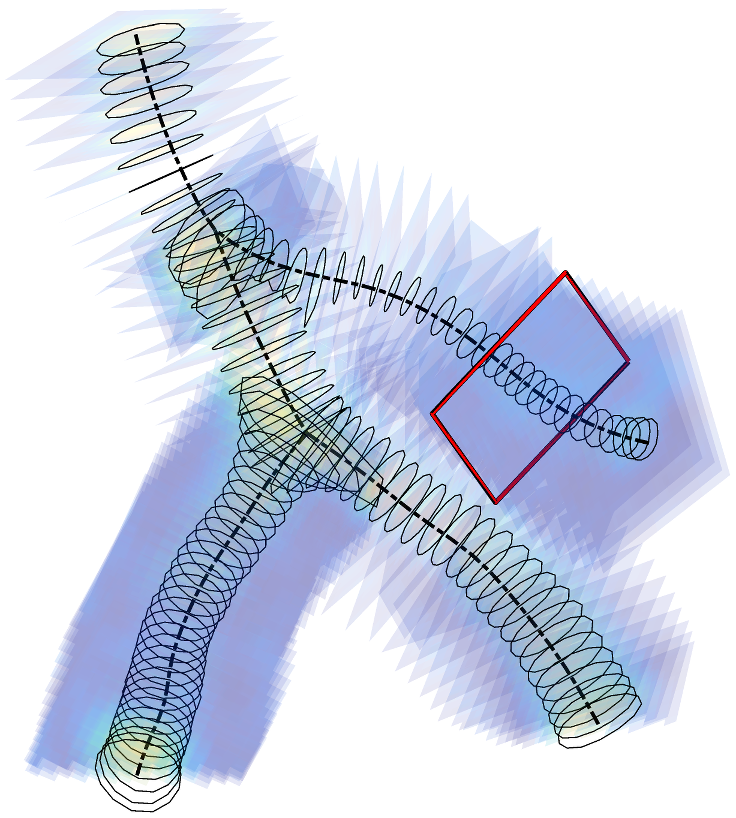}&
         \includegraphics[width=0.171\textwidth]{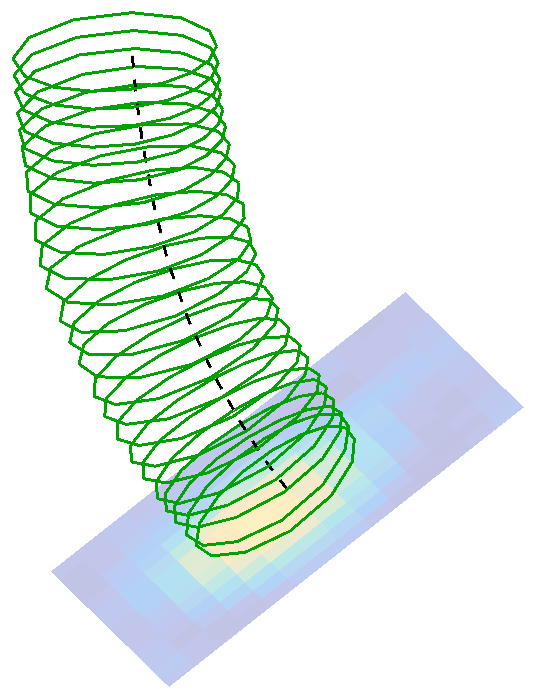}&
         \includegraphics[width=0.165\textwidth]{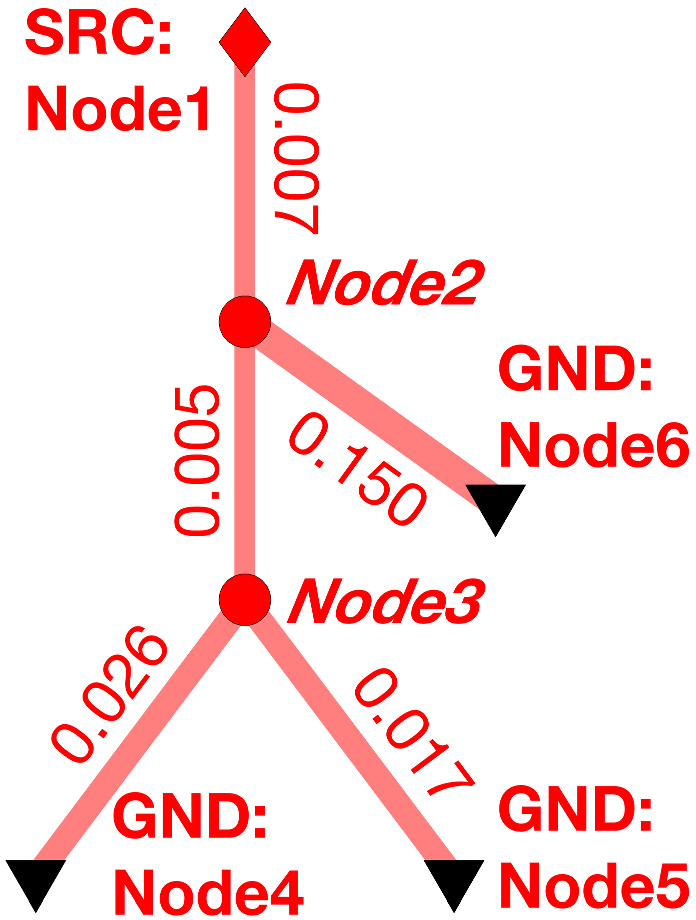}&
         \includegraphics[width=0.155\textwidth]{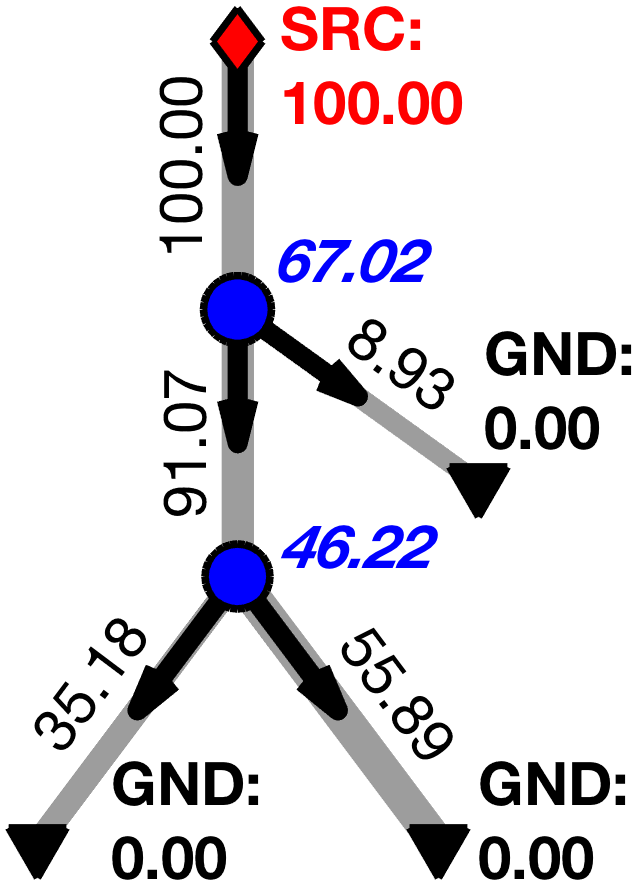}&
         \includegraphics[width=0.153\textwidth]{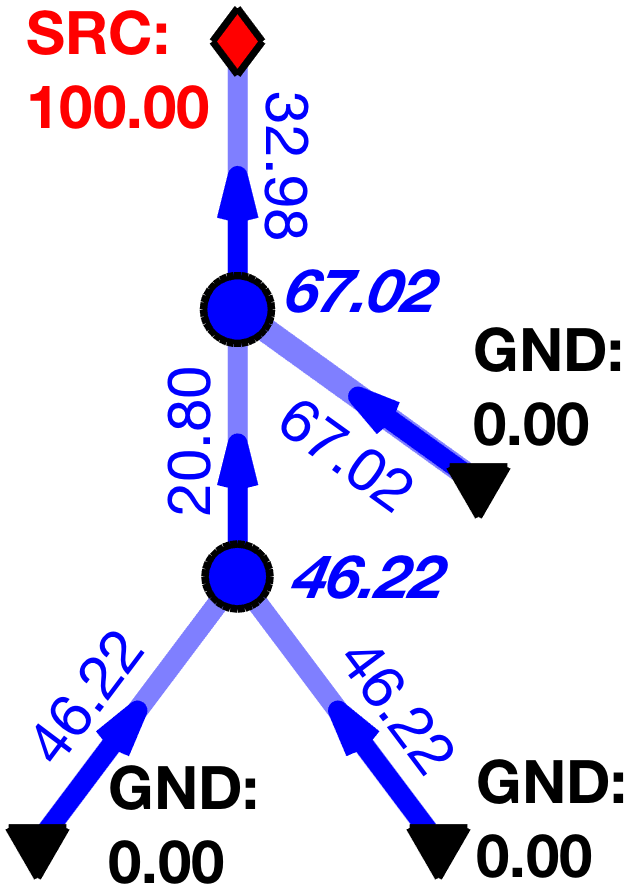}
    \end{tabular}
    \caption{Cross-sectional lumen segmentation of a vascular phantom using snakes \cite{cheng2015accurate}; hybrid vascular analog-equivalent of the phantom: impedance, flow and pressure drop.}
    \label{fig:M1}
\vskip -10pt
\end{figure}
We build on prior work by automatically extracting a vascular graph of connected centerlines from angiographic images \cite{aylward2012tubetk,VMTK2008,moriconi2018inference}, and a active-contour-based lumen segmentation model \cite{cheng2015accurate}, depicted in fig. \ref{fig:M1}.
Using this vascular representation, we first show how to create a mechanical vascular equivalent automatically from a geometrical segmentation of the neurovascular tree under the assumption of simplified hemodynamics. The extracted vascular lumped-parameter model is then converted into an analog closed-circuit configuration, enabling a computationally-efficient linear-approximation of blood flow and pressure drop. Lastly, we introduce geometrical and topological perturbations to the model to simulate how the vascular network would react to different events (e.g. embolism, stenosis) providing a measure of network resilience.

\subsubsection{Hybrid Vascular Lumped-Parameters Model}
\label{Methods:HemodynamicLumpedParametersModel}
Hemodynamic quantities are obtained from simplifying the Euler fluid equation, governing the fluid dynamics in the continuum.
Here, we approximate non-linearities and the shock of an incompressible flow transient by assuming a cylindrical model of the underlying branch geometry, where a rigid pipe runs with fixed radius along the vascular elongated (i.e. $\underline{z}$ axial) direction.
As demonstrated in \cite{vitturi2016navier}, the axial motion of a fluid is derived from the Cauchy momentum of mass conservation to the differential Hagen-Poiseuille equation, i.e. $q_{\underline{z}}^{\text{max}} = -\frac{1}{4\mu}\frac{\partial p}{\partial \underline{z}} \cdot r^{2}$, under the assumption of a steady,
fully-developed,
and axisymmetric flow $\mathbf{q}$, showing non-turbulent motion, i.e. with null flow velocity for both radial and swirl components.
The maximum flow occurs at the centre of the pipe of radius $r$,
and the constant average axial flow \mbox{$\overline{q}_{\underline{z}} = \frac{1}{2}\,q_{\underline{z}}^{\text{max}}$} integrates its parabolic profile over the pipe's cross-section.
Integrating also a linearly decreasing pressure drop $\partial p$ along the entire length $l$ of the pipe, a constant, average, axial flow $Q = \overline{q}_{\underline{z}}$ can be rewritten as 
\begin{equation}
    \label{eq10}
    Q = \frac{\Delta P}{R}, ~~~~ \text{with} ~~~~ R = \frac{8\mu l}{\pi r^4} ~~~~ \text{and} ~~~~ r = \frac{1}{l}\int_{l}\sqrt{\frac{a(\sigma(\underline{z}))}{\pi}}\,d\underline{z},
\end{equation}
with $\Delta P$ the integral pressure gradient, $R$ the average resistance of the rigid pipe of radius $r$, and $\mu$ the constant blood viscosity.
In this work, the constant radius $r$ is averaged along the pipe using the area of cross-sectional snakes $a(\sigma(\underline{z}))$.
\subsubsection{Graph-based Analog-Equivalent}
\label{Methods:0DTopologicalAnalogEquivalent}
Along with the hydraulic analogy of electric systems, we model analog-equivalent circuits as a set of connected lumped-parameters for the vascular network.
A generic vascular graph $G = (N,E)$ is defined as a set of nodes $j = 1,...,|N|$ (i.e. the branch-points), and the associated connecting edges $e_{(j_1,j_2)}$ (i.e. the vascular branches), encode in $E(j_1,j_2)$ the binary adjacency matrix.
For each $e_{(j_1,j_2)}$, the tubular features are converted into electrical impedance for an analog equivalent, where purely dynamic components vanish for a steady-state flow. The impedances of the connected pipes simplifies to real-valued resistances $R = f(l,r)$ as in \cref{eq10}.
These are embedded in the associated resistance-weighted adjacency matrix $R(j_1,j_2) = R_{e_{(j_1,j_2)}}$.
In a similar form, the flow $Q(j_1,j_2)$ and the pressure drop $\Delta P(j_1,j_2)$ are translated into current and potential difference for each vascular branch, respectively.
Simulating a closed-loop analog circuit, voltage generators ($\textit{SRC}_{j}$) and potential grounds ($\textit{GND}_{j}$) are introduced in the system. These model the pressure at the inlets or outlets of the network as node-wise potential boundary conditions ($P_{\textit{BC}}$).
By coupling linear lumped-parameters and the set of boundary conditions, the analog-equivalent circuit is solved using Kirchhoff's laws as a linear system of equations. As described in \cref{alg1}, $C_{R^{-1}}$ is the circuit admittance matrix. $C_{R^{-1}}$ is initialised with negative and inverse resistance values and subsequently integrates the equivalent topological system where $D_{R^{-1}}$ and $A_{R^{-1}}$ represent the associated diagonal degree and the adjacency matrix respectively, as in a canonical graph Laplacian. $C_{P_{\textit{BC}}}$ is the node-wise potential vector of boundary conditions, and $P$ is the node-wise potential solution of the linear system of equations. The canonical passive sign convention is enforced (\cref{fig:M1}).
\begin{algorithm}[t]
\vskip -10pt
\caption{Graph-based analog-equivalent system: definition and solver.}
\label{alg1}
\textbf{Input:} $R$, $P_{\textit{BC}}$; ~~~ \textbf{Output:} $P$, $\Delta P$, $Q$\\
$C_{R^{-1}}(j_1,j_2) = - R(j_1,j_2)^{-1}$; ~~~ $C_{P_{\textit{BC}}} = \mathbf{0}_{|N| \times 1}$; \Comment{Initialisation}\\
$C_{R^{-1}} = D_{R^{-1}} - A_{R^{-1}}$; \Comment{Circuit Admittance System}\\
\For{\textbf{all} $P_{\textit{BC}_{j}} \in \lbrace\textit{SRC},\textit{GND}\rbrace$}{
$C_{R^{-1}}(j_{1_{==}} j,\forall j_2) = 0$; ~ $C_{R^{-1}}(j_{1_{==}} j,j_{2==} j) = 1$;\\
$C_{P_{\textit{BC}}}(j_{1==} j) = P_{\textit{BC}_{j}}$; \Comment{Include Boundary Conditions}\\
}
$P = C_{R^{-1}}^{-1}\,C_{\textit{PB}}$; \Comment{Solve the Linear System}\\
\For{\textbf{all} $E(j_1,j_2)~{==}~1$}{
$\Delta P(j_1,j_2) = P(j_1) - P(j_2)$; \Comment{Assign Potential Difference}\\
$Q(j_1,j_2) = \Delta P(j_1,j_2) \cdot R_(j_1,j_2)^{-1}$; \Comment{Assign Current (Ohm's law)}\\
}
\end{algorithm}
\vspace{-10pt}
\subsubsection{Modelling Perturbations on Vascular Topologies}
\label{Methods:ModellingPerturbationsOnVascularTopologies}
\begin{figure}[b]
	\vskip -12pt
    \centering
    \begin{tabular}{cccc}
    \centering
         \tiny{\textbf{Time Of Flight MRA}}&
         \tiny{$G |_{\mathcal{A}} \cup \sigma$}&
         \tiny{\textbf{CoW Topology}}&
         \tiny{\textbf{Types}}
         \\
         \includegraphics[width=0.3\textwidth]{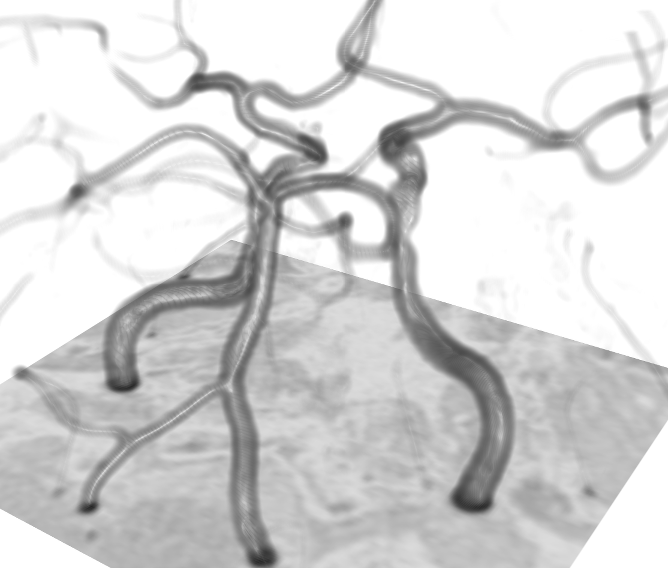}&
         \includegraphics[width=0.3\textwidth]{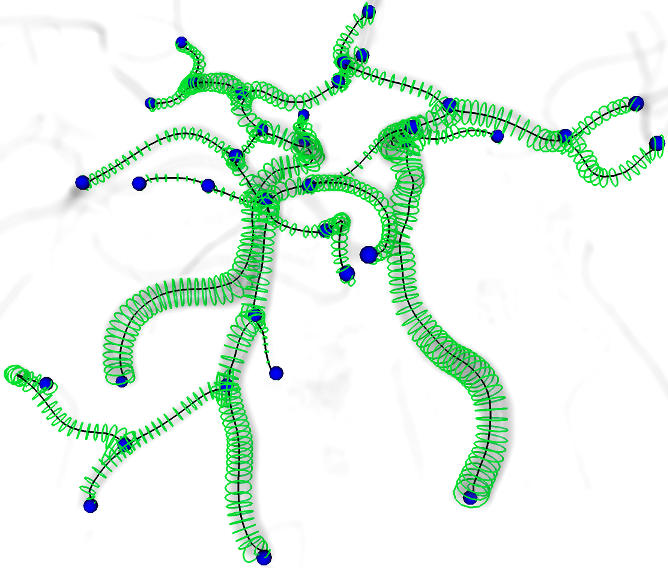}&
         \includegraphics[width=0.26\textwidth]{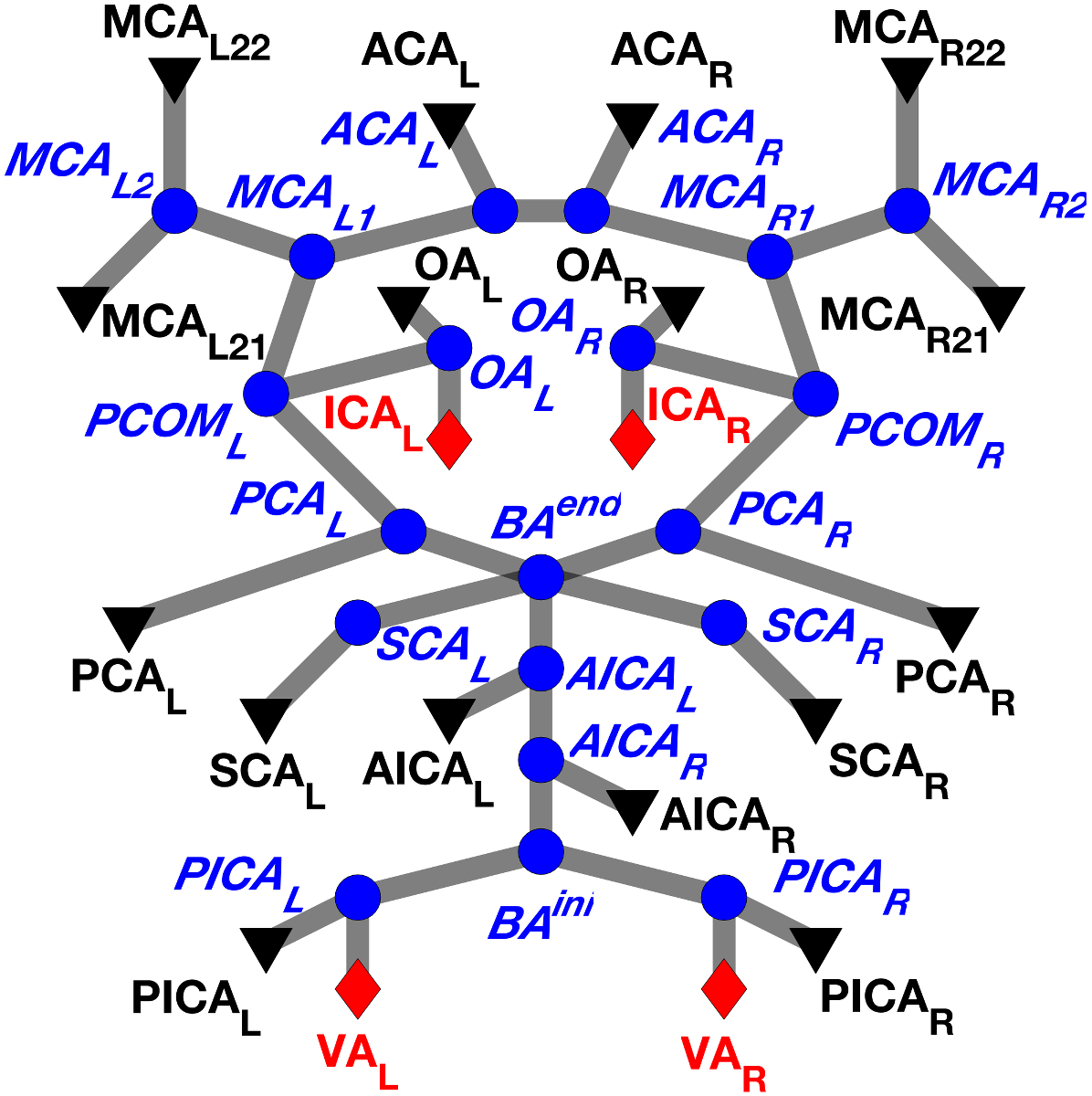}&
         \raisebox{.115\textwidth}{
         \begin{tabular}{@{}c@{}}
              \includegraphics[width=0.08\textwidth]{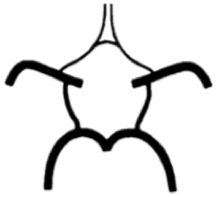}\\
              \includegraphics[width=0.08\textwidth]{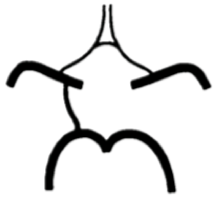}\\
              \includegraphics[width=0.08\textwidth]{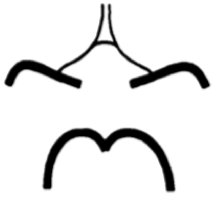}
         \end{tabular}
         }
    \end{tabular}
    \caption{Circle of Willis: MR angio, snake segmentation, manual landmarks and graph equivalent for an exact network ($\mathcal{A}$). \textit{SRC} ($\lozenge$) and \textit{GND} ($\blacktriangledown$) are shown for in/outlets.}
    \label{fig:R0}
\vskip -12pt
\end{figure}
We introduce two types of perturbations to account for changes in structural connectivity and flow resistance modulation $\mathcal{M}$. The structural connectivity perturbation is achieved by altering $\tilde{G}(N,\tilde{E})$ with a mask $E$ as $\tilde{E} = \mathcal{E} \circ ( E \circ \mathcal{A} )$, where $\mathcal{E}(j_1,j_2) \sim \mathcal{B}(\lambda)$. Here, $\mathcal{E}$ follows a Bernoulli distribution $\mathcal{B}$ of probability $\lambda$. These model random occlusions, which disrupt the connectivity by a factor \mbox{$\varepsilon = (1 - \lambda)$}, on average.
$\mathcal{A}$ is an anatomical prior where non-zero edges $\mathcal{A}(j_1,j_2)$ weight the likelihood of certain cerebrovascular connections.
In general $\mathcal{A}$ is \textit{unknown} for non-annotated graphs, meaning that $\mathcal{A} = \mathbf{1}_{|N| \times |N|}$, therefore vanishing in $\tilde{E}$.
Prior knowledge can be embedded in $\mathcal{A}$ if available.
The second type of perturbation, namely vascular stenoses and vessel tortuosity, are modelled for both reduced radii $r$ and longer pipes' lengths $l$, respectively.
These are element-wise integrated in $\tilde{R}_{\tilde{E}} = \mathcal{M} \circ R_{\tilde{E}}$, with \mbox{$\mathcal{M}(j_1,j_2) \sim  1 - U(0,m)$}, following a uniform distribution with $m < 1$. 
Note that these perturbations are computationally very efficient, as they are defined as simple matrix-to-matrix element-wise transformations.
\vspace{-6pt}
\section{Experiments and Results}
\label{ExperimentsResults}
\vspace{-6pt}
\paragraph{Datasets:}
\label{ExperimentsResults:Datasets}
We use six MR time-of-flight angiographies of the Circle of Willis (CoW), with each subject is classified into 3 different CoW phenotypes and manually labelled as in \cref{fig:R0}, and vascular graphs are extracted following \cite{moriconi2018inference}.
\vspace{-6pt}
\subsubsection{Controlled Simulations on \textit{Exact} Topologies (CoW)}
\label{ExperimentsResults:Exp2}
\begin{figure}[!b]
\vskip -8pt
    \centering
    \begin{tabular}{@{}c@{}c|c@{}c@{}}
        \raisebox{7.5em}{(a)}&
         \begin{tabular}{@{}cc@{}}
              $G(~$\raisebox{-0.5em}{\includegraphics[width=0.05\textwidth]{CoW_topoA.png}}$~)$&
              $\tilde{G}(~$\raisebox{-0.5em}{\includegraphics[width=0.05\textwidth]{CoW_topoA.png}}$~)$\\
              \includegraphics[width=0.3\textwidth]{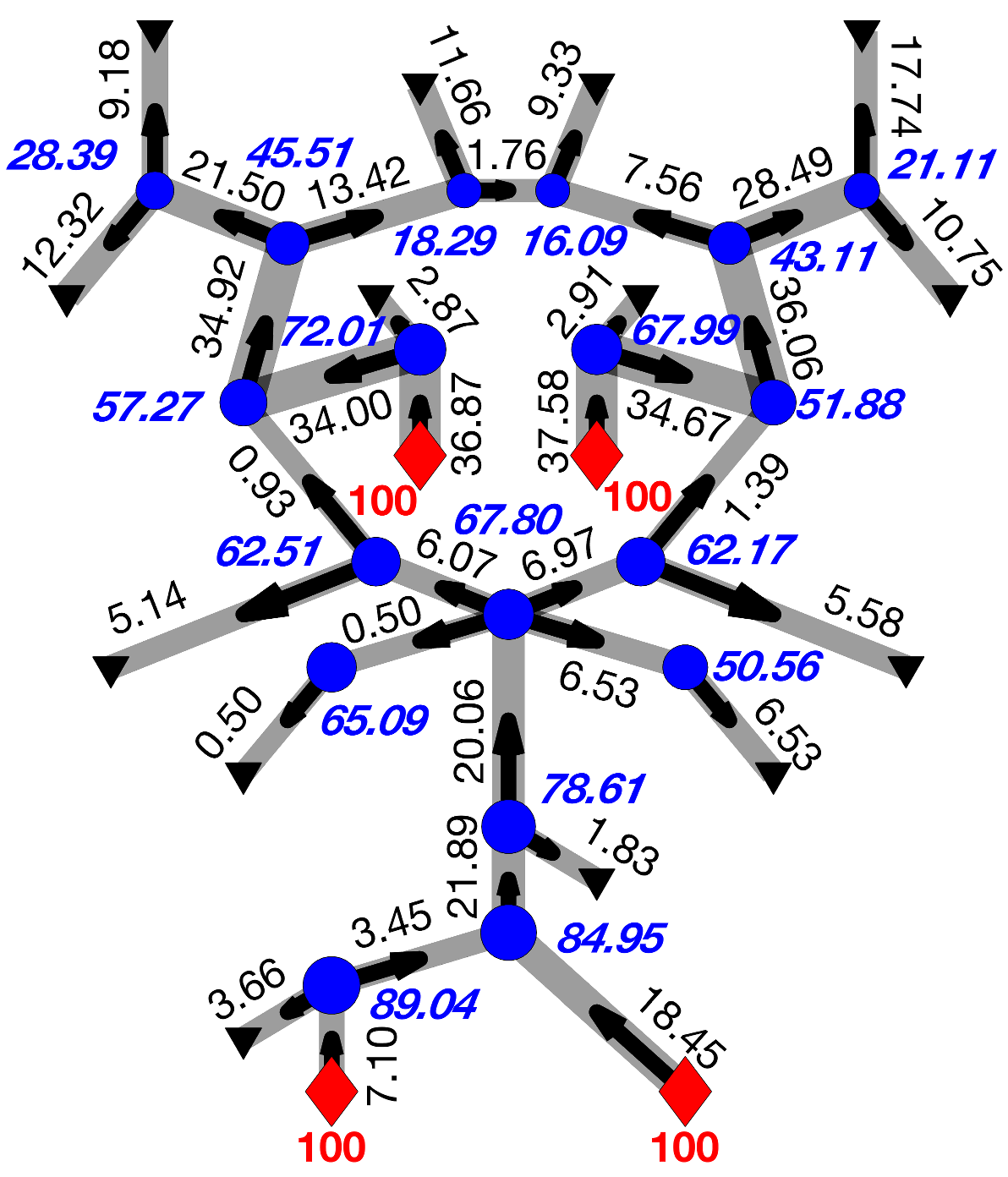}&
              \includegraphics[width=0.3\textwidth]{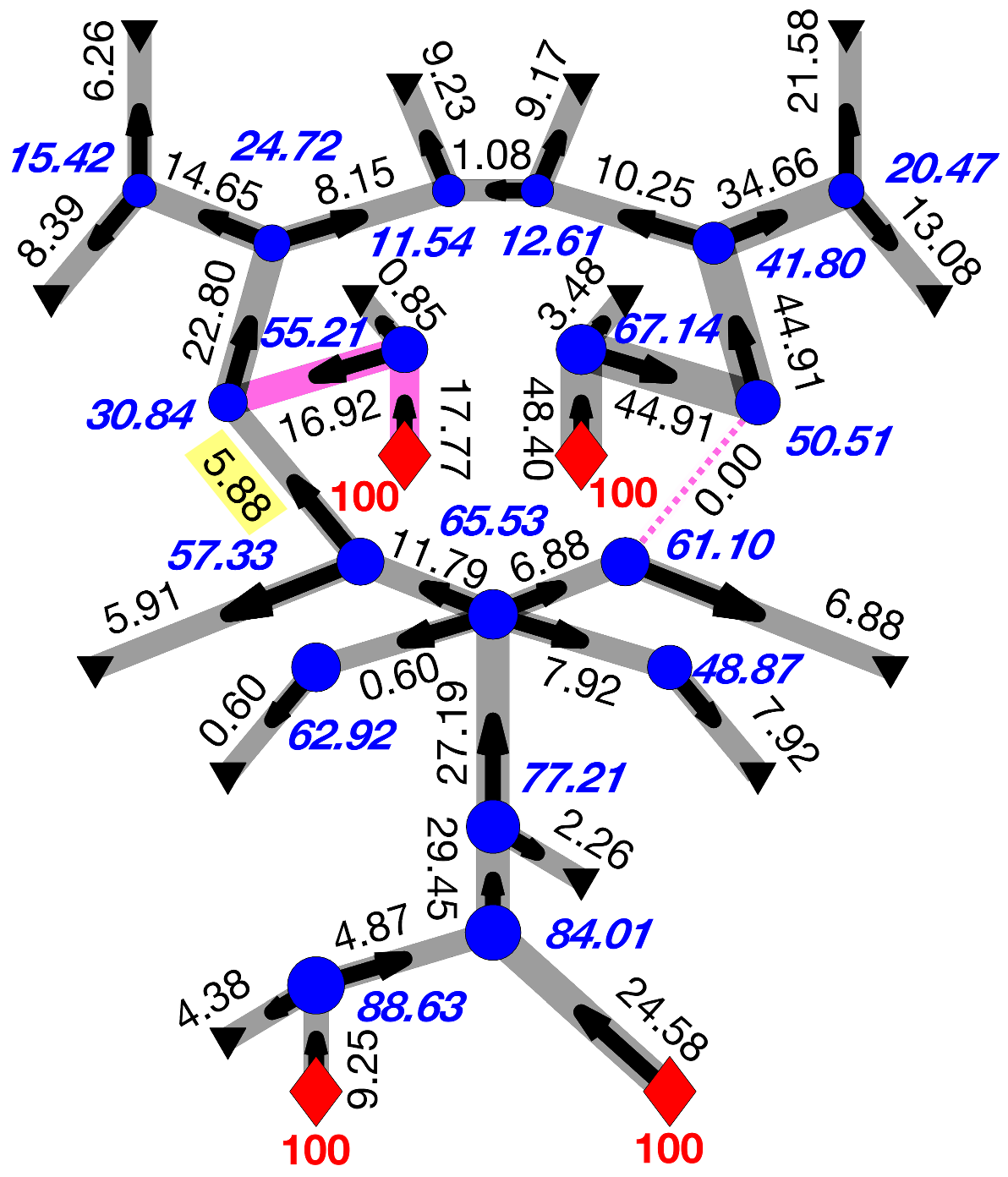}\\
              $\rho_{G} = 23.7$&
              $\rho_{\tilde{G}} = 20.8$
         \end{tabular}
         &
         \begin{tabular}{@{}cc@{}}
              $G(~$\raisebox{-0.5em}{\includegraphics[width=0.05\textwidth]{CoW_topoB.png}}$~)$&
              $\tilde{G}(~$\raisebox{-0.5em}{\includegraphics[width=0.05\textwidth]{CoW_topoB.png}}$~)$\\
              \includegraphics[width=0.14\textwidth]{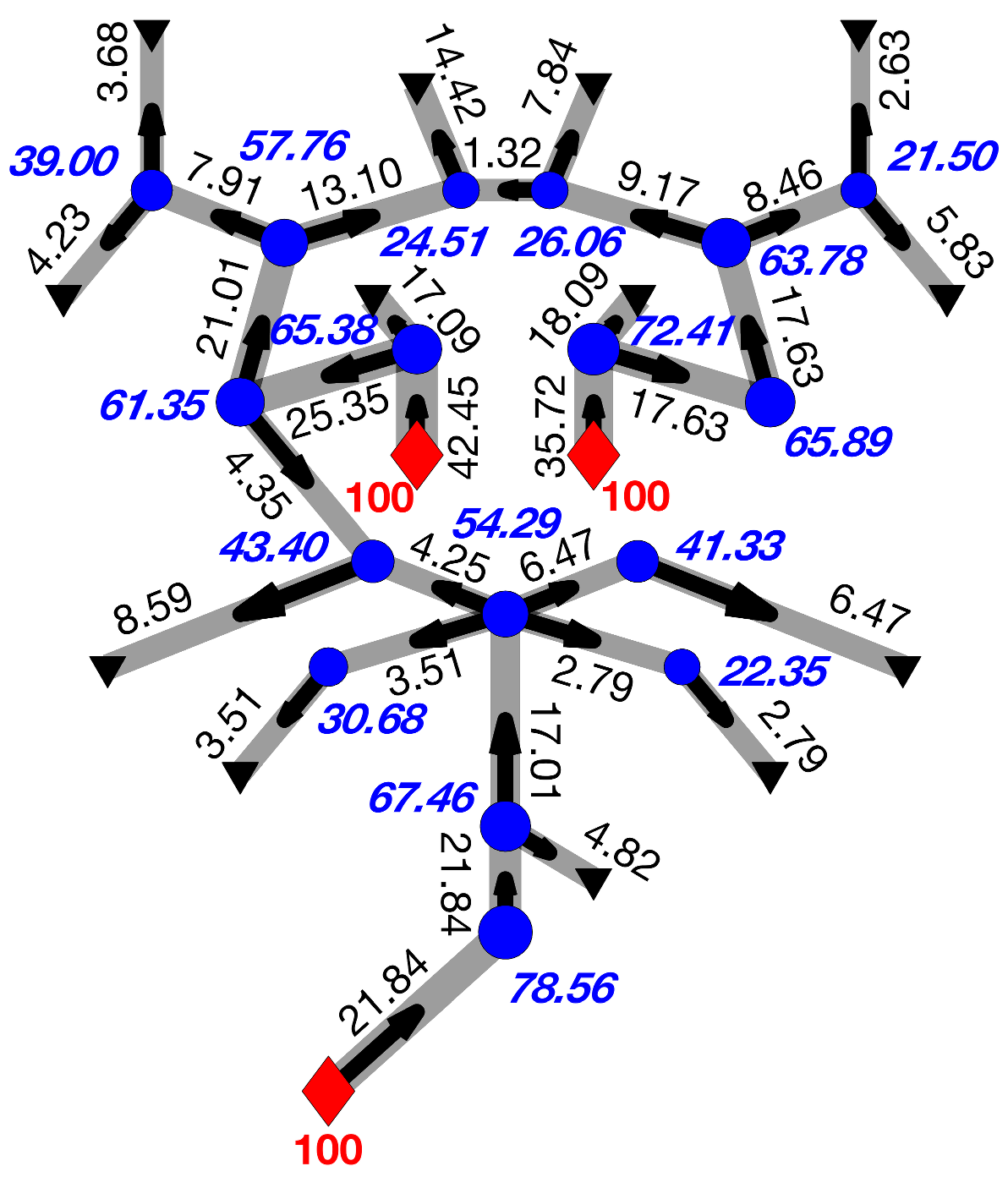}&
              \includegraphics[width=0.14\textwidth]{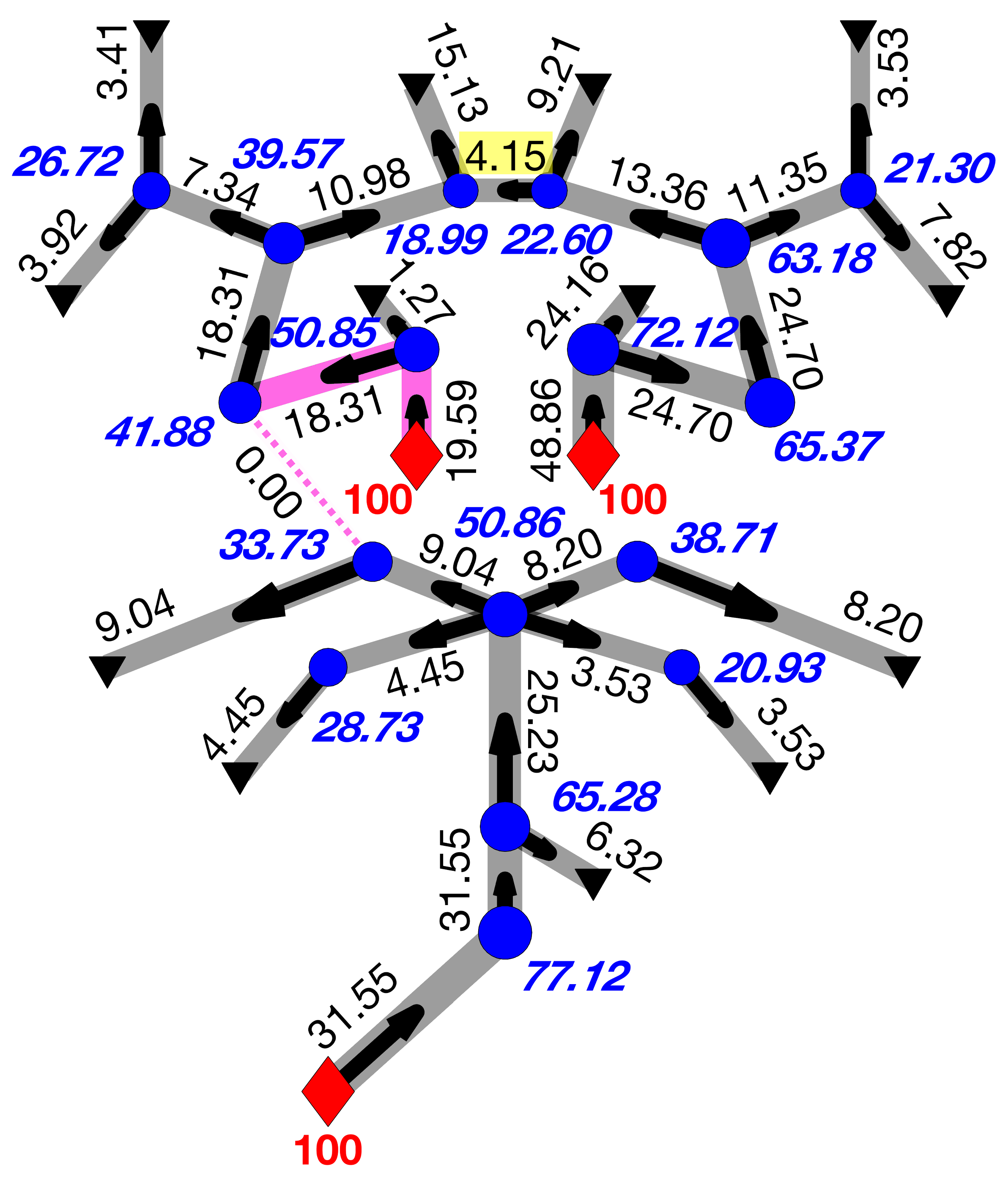}\\[-0.6em]
              \tiny{$\rho_{G} = 18.0$}&
              \tiny{$\rho_{\tilde{G}} = 16.0$}\\
              \cline{1-2}\\[-1em]
              $G(~$\raisebox{-0.5em}{\includegraphics[width=0.05\textwidth]{CoW_topoC.png}}$~)$&
              $\tilde{G}(~$\raisebox{-0.5em}{\includegraphics[width=0.05\textwidth]{CoW_topoC.png}}$~)$\\
              \includegraphics[width=0.14\textwidth]{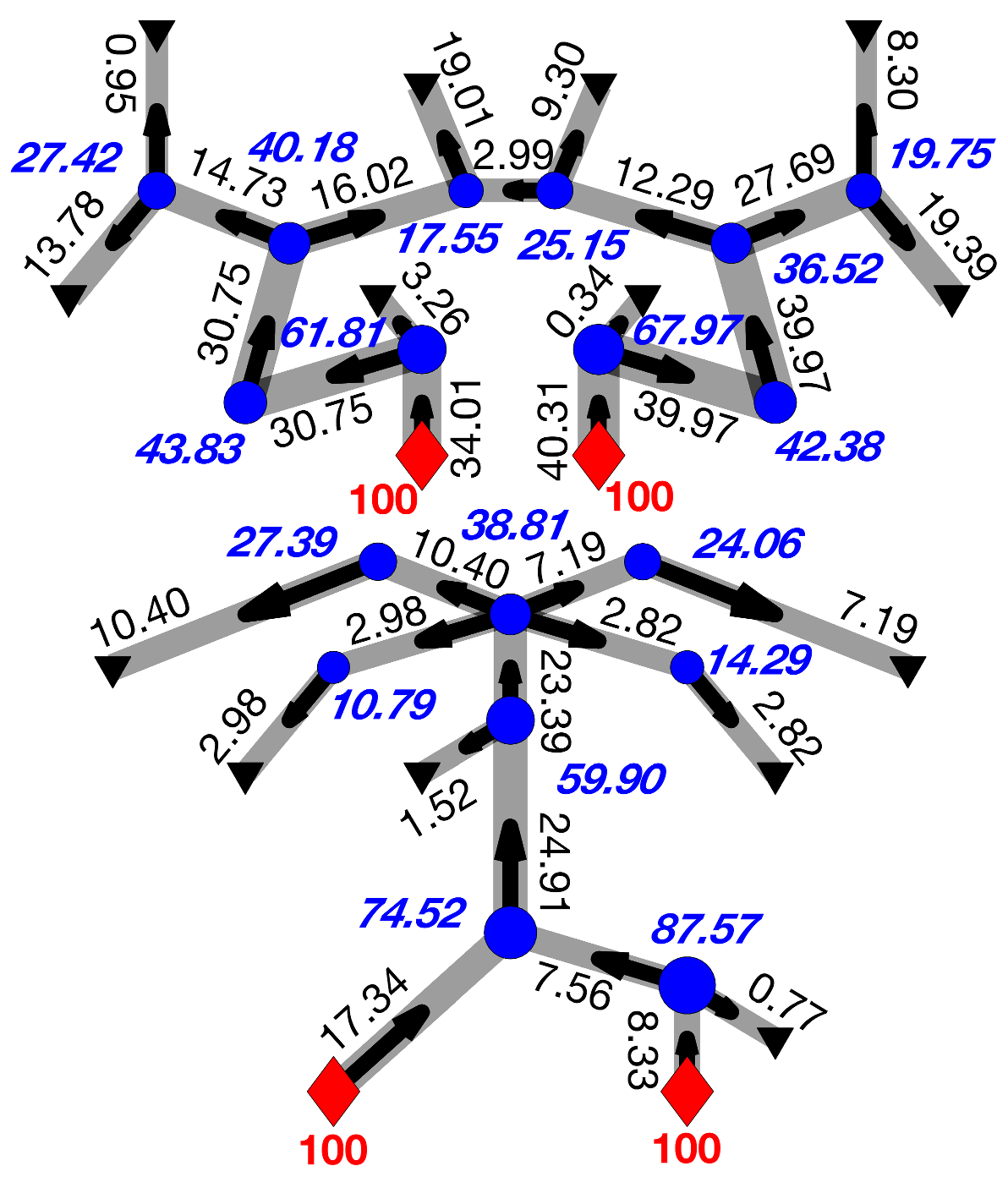}&
              \includegraphics[width=0.14\textwidth]{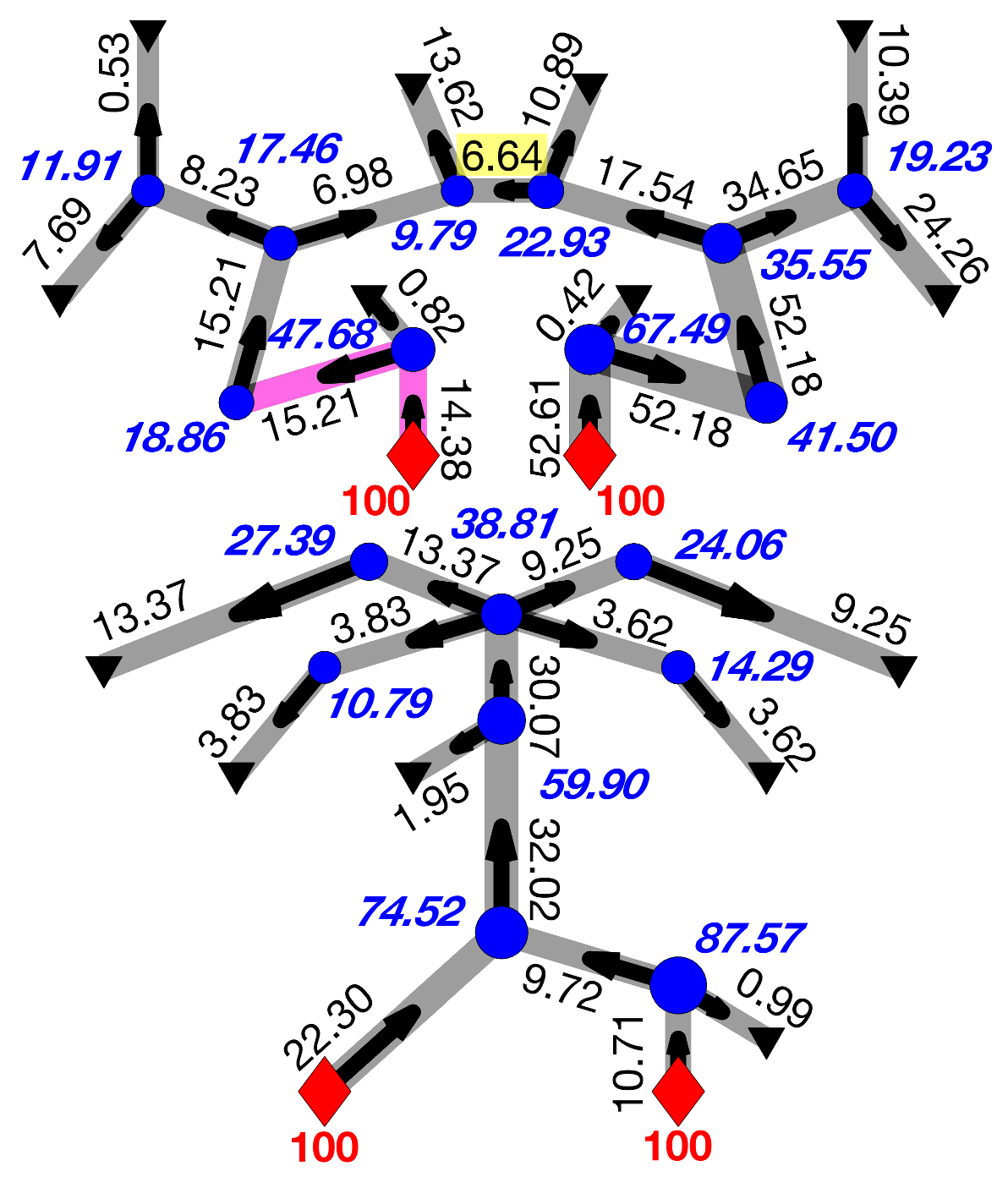}\\[-0.6em]
              \tiny{$\rho_{G} = 17.0$}&
              \tiny{$\rho_{\tilde{G}} = 15.5$}
         \end{tabular}
         &\raisebox{7.5em}{(b)}\\[-0.95em]
         &&&(c)
    \end{tabular}
    \caption{Autoregulation mechanisms: blood flow, pressure and network resilience $\rho$ for unperturbed graphs, and for the simulated stenotic \small{ICA} \normalsize and occluded \small{PCOM}\normalsize.}
\vskip -12pt
    \label{fig:R2}
\end{figure}
Pressure potentials are initialised at the anatomical inlets, whereas potential grounds are set at the terminal branches of the CoW (\cref{fig:R0}).
Given an anatomical prior $\mathcal{A}$ from the annotated graph, we first evaluate the autoregulation mechanisms by simulating a stenotic Internal Carotid Artery (\small{ICA}\normalsize) and an occlusion of the Posterior Communicating Artery (\small{PCOM}\normalsize), as sanity test in a simple, yet realistic, scenario.
In \cref{fig:R2} biologically compatible autoregulation mechanisms are observed for different types of the Circle of Willis.
On average, reduced flow and pressure values are found in the perturbed ipsi-lateral branch of the network, whereas minimally affected quantities are observed for the contra-lateral part.
While flow is marginal in the PCOMs for the unperturbed network \cref{fig:R2} (a), it increases (highlight) after the simulated stenosis and occlusion (purple edges), where the flow overdraft is compensated by the posterior circulation.
Similar autoregulation mechanisms are observed for the other phenotypes \cref{fig:R2} (b) and (c), where major compensations are given by the anterior left-right circulatory contribution at the Anterior Communicating Artery (\small{ACA}\normalsize) level. Flow readjustment were intrinsically different for different CoW phenotypes. 
Despite the relatively small \small{ACA} \normalsize size, increased flow is observed (highlighted) contro-lateral to the simulated perturbation.
Lastly, resilience indices $\rho$ show a decreasing trend for the \textit{same} perturbation on the different networks.
\vspace{-8pt}
\subsubsection{Random Perturbations on \textit{Exact} Topologies (CoW)}
\begin{figure}[!b]
\vspace{-12pt}
    \centering
    \begin{tabular}{@{}c@{}cc|@{}c@{}}
         \tiny{\textbf{Type}}&\tiny{\textbf{Blood Flow} $Q$}&\tiny{\textbf{Blood Pressure} $P$} & \tiny{\textbf{ Hypertension Histogram}} \\
         \raisebox{2.5cm}{
         \begin{tabular}{c}
            \includegraphics[width=0.05\textwidth]{CoW_topoA.png}\\\\[0.5em]
            \includegraphics[width=0.05\textwidth]{CoW_topoB.png}\\\\[0.5em]
            \includegraphics[width=0.05\textwidth]{CoW_topoC.png}
         \end{tabular}
         }&
         \includegraphics[width=0.3\textwidth]{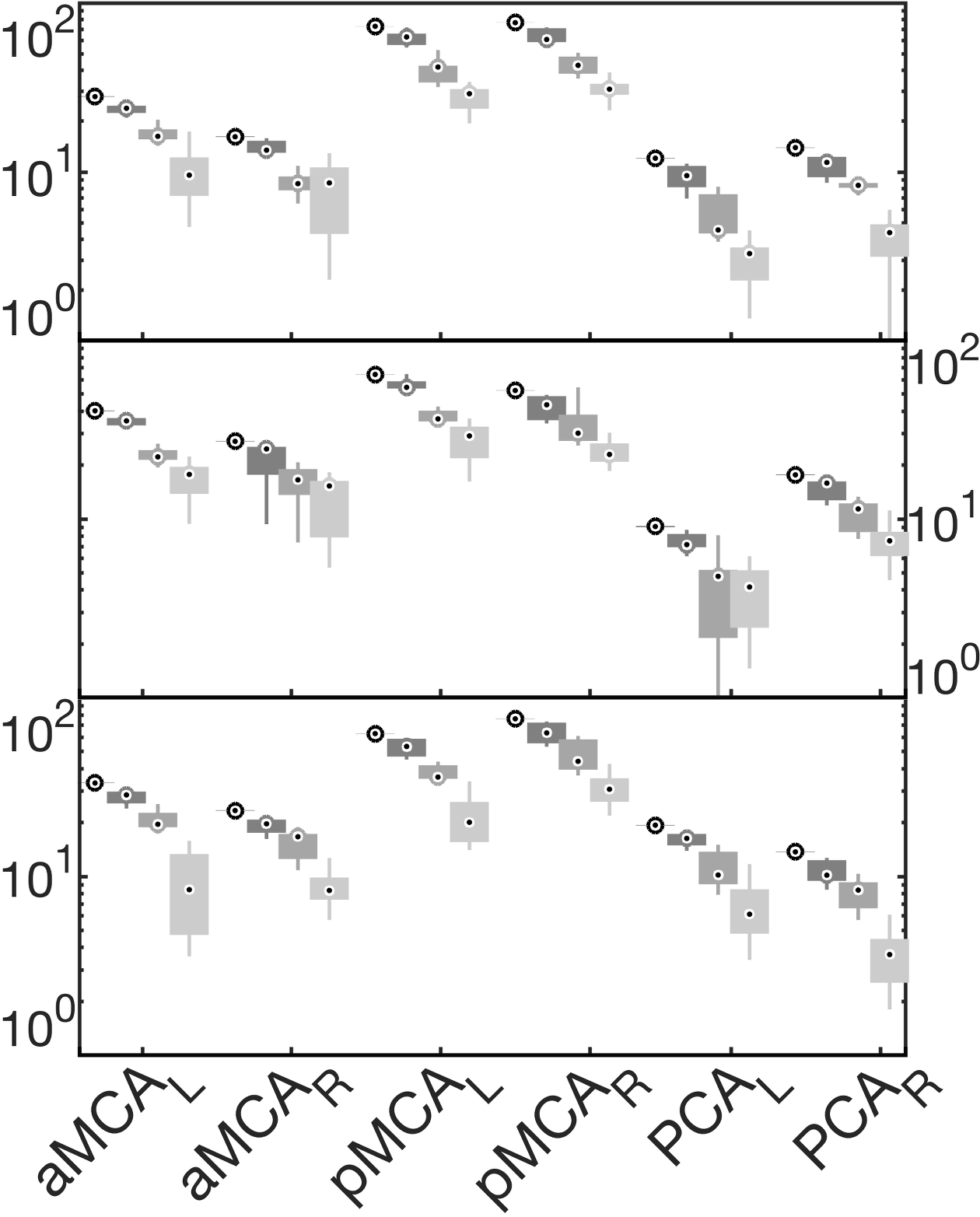}&
         \includegraphics[width=0.32\textwidth]{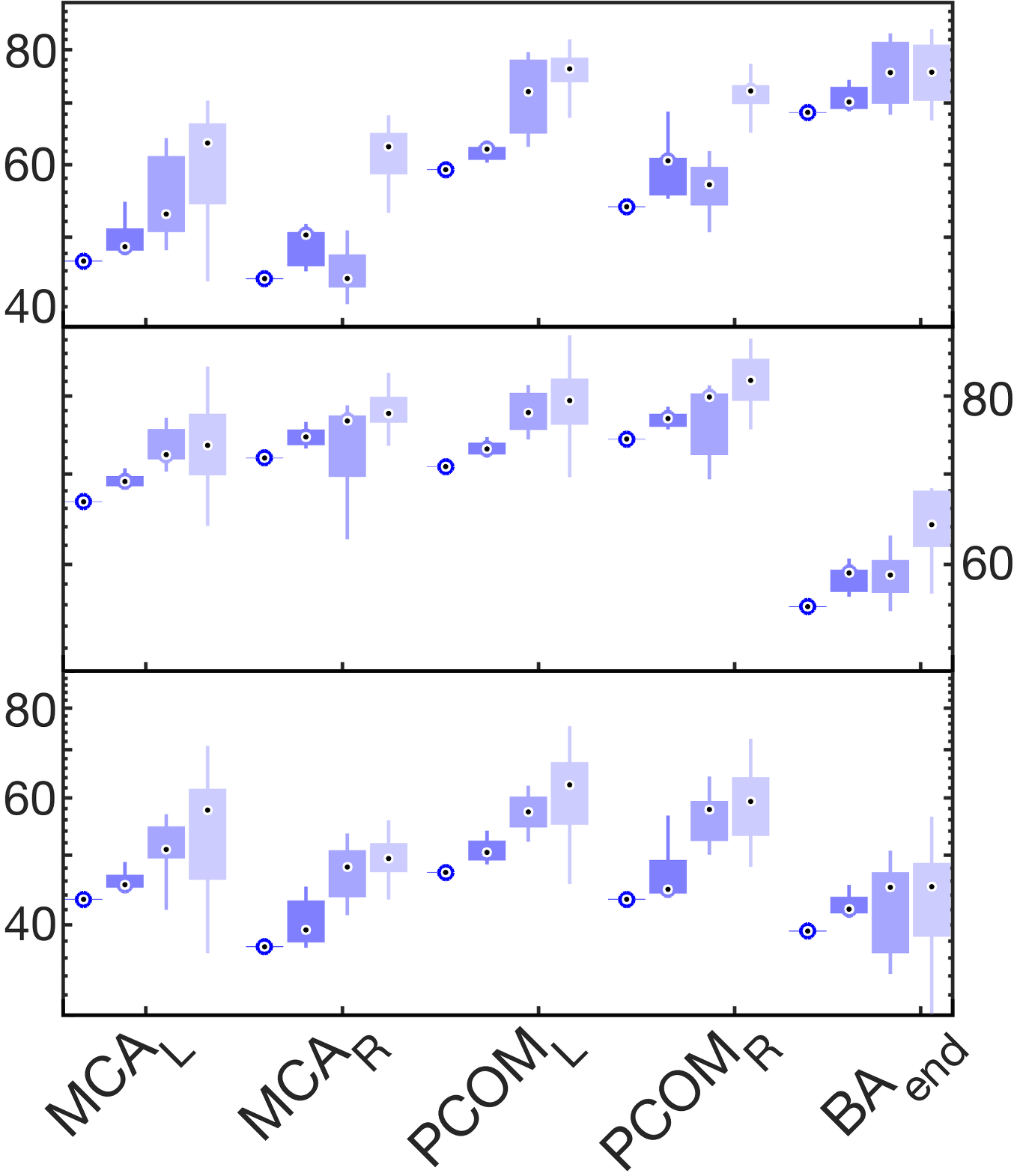}&
         \raisebox{1.5em}{
         \includegraphics[width=0.3\textwidth]{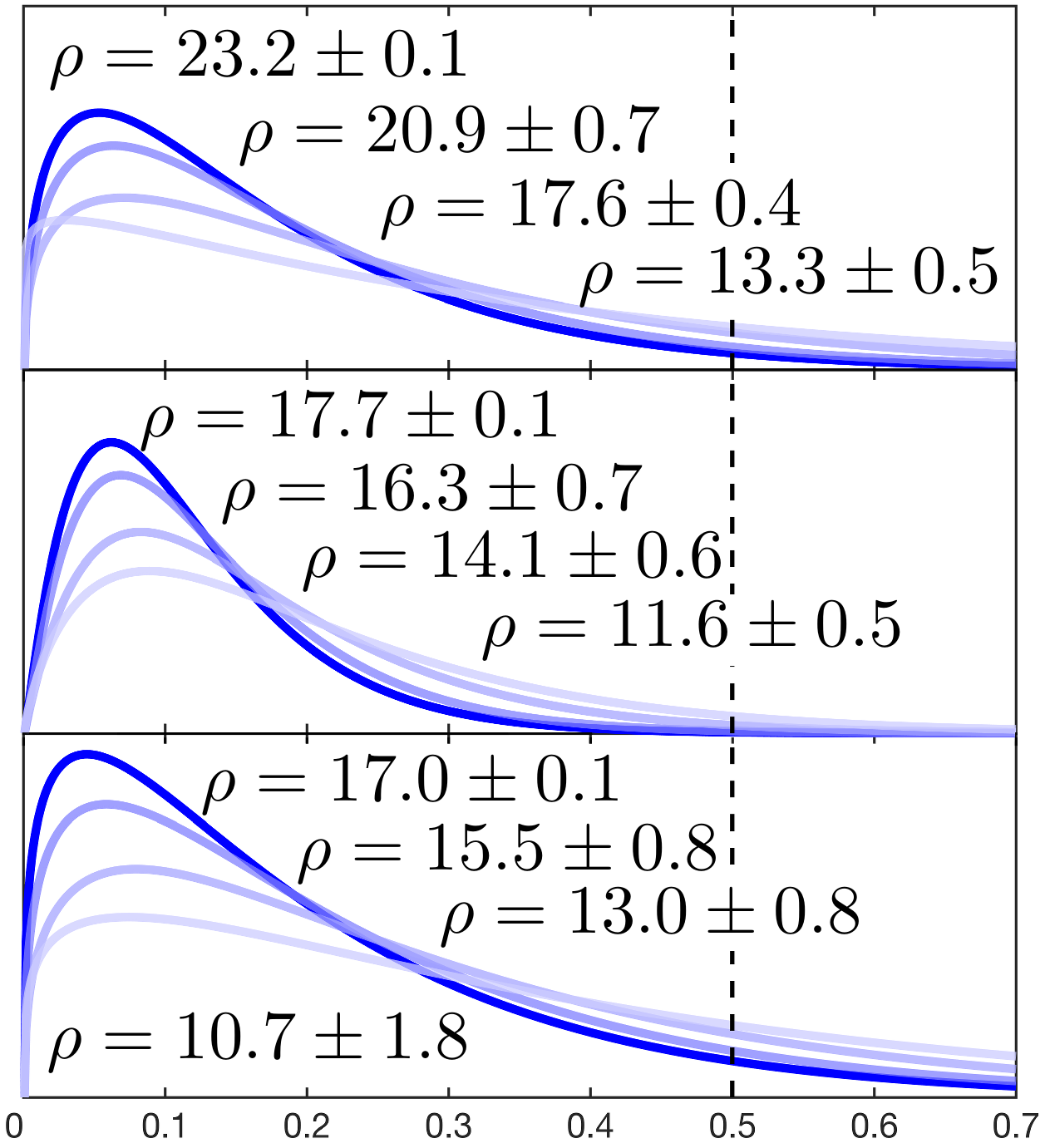}}
    \end{tabular}
    \caption{Flow, pressure and hypertension histogram for unperturbed anatomically exact topologies and for 3 perturbation classes of increasing stenoses and tortuosity.}
    \label{fig:R3}
\vspace{-14pt}
\end{figure}
Simulations are computed by perturbing \textit{only} the resistance-equivalents, where fluctuations in $\mathcal{M}$ modulate  both radius and length of the pipes.
Perturbations account for 3 classes with maximal resistance increment $m^{\text{max}} = 50\%$, and a total $n = 1000$ instances per class.
Flow and pressure are evaluated also on unperturbed equivalents for comparison.
The $n$ simulated quantities are then averaged for the same phenotype.
A global network resilience metric is defined as $\rho_{G} = \frac{1}{|\tilde{E}|}  \sum_{|\tilde{E}_{}d|} \rho \, $, with $\rho = \frac{\left( \Delta P \cdot Q \right)}{ \pi r^2 l } \, $, where $|\tilde{E}|$ is the total number of vascular branches \textit{after} the topological perturbation (here $\mathcal{E} = \mathbf{1}_{|N| \times |N|}$), being $|\tilde{E}_{d}|$ the diffused ones, i.e. those having non-zero $Q$ and $\Delta P$.
The scalar $\rho_{G}$ is an integral surrogate for the branch functionality given the network perturbation. In order words, it is a scalar describing how resilient, or how affected, a branch is by a random perturbation anywhere in the neurovascular tree. 
Assuming healthy networks being well diffused, $\rho_{G}$ is maximal for unperturbed equivalents, whereas it decreases for impairing modulations.
In \cref{fig:R3}, flow and pressure distributions are depicted for a representative set of CoW edges and nodes.
For each perturbation class (i.e 0$<$$m_1$$<$0.2, 0.2$<$$m_2$$<$0.3, and 0.3$<$$m_3$$<$0.5), the hemodynamic quantities are compared against the unperturbed values.
On average, flow is decreased, in line with the overall increased impedance of the vascular network.
Conversely, the distributed pressure increases progressively as the degree of perturbation, with relatively smaller ratio of increase at the Basilar Artery terminal point (\small{$\text{BA}^{\text{end}}$}\normalsize).
\indent As a second tier analysis, a hypertension histogram is fitted in \cref{fig:R3} with a gamma distribution.
Here, hypertension is  defined as the pressure normalised by the mean cross-sectional area of the vessel.
An unperturbed CoW shows a hypertension profile skewed towards low values;  histograms shows a broader profile for increasing perturbations,  with more small vessels reporting relatively high pressure.
This suggests that increased hypertension tends to affect the whole CoW even for localised stenoses (leftwards shift of the histogram), and increased risk of vascular rupture (larger area above a certain threshold, e.g. the dashed line in fig \ref{fig:R3}).
A higher prevalence of zero-force is also observed in the simulated stenotic regions, suggesting higher risk of ischemia. Resilience $\rho$ was also found to decrease for all topologies at increasing levels of perturbation.
\vspace{-8pt}
\subsubsection{Perturbations on Redundant Uncertain Topologies (CoW)}
\label{ExperimentsResults:Exp3}
%
%
So far, analyses assumed a specific realisation of a vascular graph.
However, robustly extracting the vascular topology is a challenging task due to poor image resolution.
In our validation, we observed that erroneously extracted vascular graphs, i.e. those with the wrong connections, exhibited abnormal biomechanical properties.
We thus hypothesise that the proposed simulation framework can assess the plausibility of putative vascular graphs.
Relaxing the assumption of a known anatomical prior (i.e. vanishing $\mathcal{A}$), occlusive perturbations $\mathcal{E}$ are introduced for an over-connected graph $G$, which embeds uncertainty among vascular junctions.
Similar boundary conditions are initialised for those nodes closest to the annotated in/outlets; however, no resistance modulation is performed.
Note that simulating complete occlusions on a redundant vascular network is equivalent to re-sample $G$ with subnets and evaluate their biological compatibility.
Here, three classes of randomly occluded topologies $\tilde{G}$ are generated for $\varepsilon = 0.2,\,0.3,\,0.5$, each with a total of $n = 1000$ instances.
For each class, an inverse resilience adjacency matrix $\hat{\rho}$, of the same size as $\tilde{E}$, is determined as $\hat{\rho}(j_1,j_2) = {\rho(j_1,j_2)}^{-1}$, and an associated likelihood matrix $\mathcal{L}$ is integrated for all simulations in each class.
Specifically, $\mathcal{L} = \sum_{n} \rho_{n} \cdot \textit{\text{MST}}(\hat{\rho_{n}})$, where \textit{MST} is the minimum spanning tree maximising the resilience of each perturbed instance.

From our experiments we obtain putative re-sampled graphs $\tilde{G}$ resulting in subsets of most hemodynamically-compatible branches from an initial fully-connected topology.
Major sparsity in the associated adjacency matrix is found for $\varepsilon = 0.5$ and by thresholding the likelihood $\mathcal{L}$ above the median.
Although the supra-threshold $\tilde{G}$ shows a reduced redundancy in the connectivity pattern, the correct CoW phenotype is kept \textit{intact}.
Also, similar patterns are found for $\varepsilon = 0.2,\,0.3$.
This suggests that for $n_{\xrightarrow[]{\infty}}$ simulations and for different degrees of perturbations, a family of hemodynamically-compatible graphs statistically emerges from an uncertain and redundant graph, by jointly maximising the subnet resilience and by integrating overlapping minimal acyclic realisations.
\vspace{-7pt}
\section{Discussion and Conclusions}
\label{DiscussionConclusions}
\vspace{-3pt}
We present a simplified graph-based simulation framework, which statistically estimates biomarkers from a series of perturbations on the neurovascular network.
Asymptotic flow and pressure are determined from data-driven, subject-specific lumped equivalents from clinical angiographies, leveraging an analog configuration and modelling pathological conditions.
The adopted approximation cannot model vascular fluid-structure interactions, nor the effect of a pulsating flow as in fully-resolved CFD simulations.
However, the high-throughput ($0.4 \pm 0.2$ ms per simulation), the arbitrary graph scalability, and the flexibility for network perturbation allow an early evaluation of the steady-state mechanisms underlying the cerebral autoregulation in a compact and reproducible way.
For three healthy CoW phenotypes autoregulation mechanisms and functional distributions are first evaluated with a controlled perturbation, then with a series of random morphological modulations spanning over all the vascular network.
Data-driven results on exact topologies are in line with the literature, where similar compensation strategies and distributions were observed in case-studies and on artificial physio-pathological models \cite{ryu2015coupled,chnafa2017improved,onaizah2017model}.
A putative graph sampling is formulated for uncertain redundant topologies, where 
a family of compatible graphs statistically emerge from jointly maximising the subnet resilience and integrating overlapping minimal spanning trees.
Notwithstanding the novelty of presented results, which 
are the first for image-based simulations of clinically relevant neurovascular networks with perturbations, a more extensive validation is still required.
Developments will address more 
phenotypes, together with a longitudinal cohort of patients 
to evaluate the predictors 
vs. the clinical outcomes.
Also, by relaxing the steady-flow, 
time-resolved analyses will account for coupling dynamic 
modalities (e.g. arterial spin labelling) and pulsating 
simulations.
\vspace{-12pt}

\end{document}